\documentstyle[graphicx,multicol,prl,aps,epsf]{revtex}
\begin{document}
\tightenlines
\title{Quantum Dot between Two Superconductors}
\author{Yshai Avishai$^{1}$, Anatoly Golub$^1$ and Andrei D. Zaikin$^{2,3}$ }
\address{$^1$ Department of Physics, Ben-Gurion University of the Negev,
Beer-Sheva, Israel\\
$^2$ Forschungszentrum Karlsruhe, Institut f\"ur Nanotechnologie,
76021 Karlsruhe, Germany\\
$^3$ I.E.Tamm Department of Theoretical Physics, P.N.Lebedev
Physics Institute, 117924 Moscow, Russia}

\maketitle

\begin{abstract}
Novel effects emerge from an interplay between 
multiple Andreev reflections and Coulomb interaction
in quantum dot 
coupled to superconducting leads 
and subject to a finite potential bias $V$.
Combining an intuitive physical picture 
with rigorous path integral formalism we
evaluate the current $I$ through the dot and find that
the interaction shifts the subharmonic pattern of 
the $I-V$ curve is shifted
toward higher $V$. For sufficiently 
strong interaction the subgap current (at $eV < 2 \Delta$)
is virtually suppressed.
\end{abstract}
\begin{multicols}{2}
\narrowtext
Recent progress in nanotechnology 
enables the fabrication and experimental
investigation of superconducting 
contacts of atomic size with
few conducting channels \cite{Scheer,Ru}.
Transport properties of such systems are essentially determined by 
the mechanism of multiple Andreev reflections \cite{BTK}
(MAR) which is responsible for Josephson current as well as for 
dissipative currents at subgap voltages. 
Theoretical analysis of MAR and
current-voltage characteristics in small superconducting junctions
is reported in a number of papers \cite{Arnold,GZ,many}.  
In these works,
an essential ingredient is the
assumption that electron-electron interaction 
inside the contact can be neglected.
It might indeed be justified provided a metallic contact is
sufficiently large and/or strongly coupled to massive superconducting
leads. 

However, in very small contacts (quantum dots), the
Coulomb interaction is not effectively screened, 
hence it is expected to substantially affect 
transport properties of the system. 
For instance, it is well known both from theory \cite{glaz} 
and experiment \cite{RBT} that tunneling through a quantum 
dot between superconductors
can virtually be suppressed due to Coulomb effects. 
Thus, to the 
fascinating physics of SNS and SIS junctions, 
one should add that of
an SAS junction composed of superconducting leads coupled by 
an {\em interacting} quantum dot.

In the present work, the physics of 
interplay between MAR and interaction effects in 
SAS junctions subject to a finite bias is exposed.
It encodes the salient features of superconductivity, 
strong correlations and non-linear response. A 
simple intuitive physical picture is combined 
with a rigorous path integral 
technique by which
irrelevant degrees of freedom are eliminated and
an effective action is constructed (in the spirit of 
Feynman-Vernon influence functional \cite{FV}).
Similar ideas proved to be useful elsewhere,
see e.g. \cite{CL,SZ}. In the present context they have 
been applied for the relatively simple case of an $SAS$ junction 
at {\em zero} bias, focusing on the equilibrium Josephson current 
\cite{Arovas,AG}.
Our main achievements are: a) Derivation of a tractable
expression for the non-linear
tunneling current in the presence of interactions. b) 
prediction of novel physical effects pertaining to
the $I-V$ curve of an SAS junction at sub-gap bias. 

Let us then commence with a simple and physically transparent picture
of an interplay between MAR and Coulomb effects. 
Consider a quasiparticle (hole) which suffers $n$
Andreev reflections inside the superconducting junction, thereby
gaining an energy $neV$, where $V$ is the voltage bias.
As soon as $nev=2\Delta$ the
quasiparticle leaves the junction and does not contribute anymore to the
subgap current. Hence, the number of Andreev reflections $n$ for a given
voltage is $n \simeq 2\Delta /eV$.  

Assume now that the Coulomb interaction 
inside the junction is switched on. For our qualitative
discussion it suffices to account for it in terms of an
effective capacitance $C$ and its related 
charging energy $E_C=e^2/2C$. At $T=0$
and for $eV \leq E_C$ a single 
electron tunneling (and, hence, also MAR) is blocked, 
so in what follows we will consider the case
$eV>E_{C}>0$.
In order to leave the junction the quasiparticle should gain an energy 
$neV$ equal to $2\Delta +(n+1)E_C$. The last term originates from the fact
that during the MAR cycle with a given $n$ 
the charge $(n+1)e$ is transferred 
between the electrodes. Hence, an additional energy $(n+1)E_C$ should be
paid. The above condition immediately 
fixes the number $n$ at a given voltage: 
\begin{equation}
n=\left [\frac{2\Delta +E_C}{eV-E_C} \right ].
\label{n}
\end{equation}
Thus, in the presence of 
Coulomb interaction quasiparticles spend more
time inside the junction and suffer more Andreev reflections. Another
obvious observation is that at low temperature
$T$, the transfer of the charge $(n+1)e$
is blocked by interaction at voltages $eV \leq (n+1)E_C$.
Combining this observation with (\ref{n}) one arrives at the condition
\begin{equation}
eV \leq eV_{th}=E_C\left(1+\sqrt{1+\frac{2\Delta}{E_C}}\right),
\label{threshold}
\end{equation}
under which the MAR current is suppressed due to Coulomb repulsion. 
For $E_C \ll \Delta$ the voltage threshold is $eV_{th} 
\simeq \sqrt{2\Delta E_C} \gg E_C$, i.e. in this case MAR should be blocked
even at voltages {\it much higher} than $E_C/e$. For $eV_{th}\geq 2\Delta$ 
eq. (\ref{threshold}) yields
$$
E_C \geq 2\Delta /3.
$$
For such values of $E_C$ one expects the subgap current to be fully suppressed 
due to Coulomb interaction.

Let us now recall that the subharmonic peaks occur on the $I-V$ curves 
each time the MAR cycle with a given $n$ becomes impossible. Without interaction
these peaks are located at voltages $V_n=2\Delta /en$. It follows immediately
from the above discussion that
in the presence of Coulomb interaction the peaks should be shifted to
higher voltages. From eq. (\ref{n}) one finds 
\begin{equation}
V_n=\frac{E_C}{e}+\frac{2\Delta +E_C}{en},
\label{peaksEc}
\end{equation}
i.e. one expects the subharmonic peaks to be shifted by $\delta V_n=E_C(1+1/n)$
towards larger $V$ as compared to the noninteracting case. 

Thus, already a naive analysis of the interplay between MAR and Coulomb
effects allows one to predict several novel effects which can be
experimentally tested. To put these qualitative arguments 
on a firm basis we formulate 
below a realistic model of an SAS junction,
and proceed with a rigorous calculation of the $I-V$ characteristics.

{\it The model and basic formalism}. Consider, in two dimensions, 
a quantum dot at ${\bf r}=0$ weakly 
coupled to (half planar) superconducting electrodes. 
The Hamiltonian of the system is decomposed as,
\begin{equation}
\bbox{H}=\bbox{H}_{L}+\bbox{H}_{R}+ \bbox{H}_{\rm dot}+ 
\bbox{H}_{\rm t}.
\label{tot}
\end{equation}
The Hamiltonians of the left ($x<0$) and right ($x>0$) superconducting 
electrodes have the standard BCS form
\begin{eqnarray}
\bbox{H}_j=\int d\bbox{r} [\Psi^{\dagger}_{j \sigma}
(\bbox{r})\xi({\bf \nabla})\Psi_{j \sigma}(\bbox{r}) 
\nonumber \\  
	-\lambda \Psi^{\dagger}_{j \uparrow}(\bbox{r})\Psi^{\dagger}_{j \downarrow}(\bbox{r})
	\Psi_{j \downarrow}(\bbox{r})\Psi_{j \uparrow}(\bbox{r})].
\label{BCS}
\end{eqnarray}
Here $\Psi^{\dagger}_{j \sigma}$ ($\Psi_{j \sigma}$) 
are the electron creation
(annihilation) operators, $\xi ({\bf \nabla})=-{\bf \nabla}^2/2m-\mu$, 
and $j=L,R$ for left and right electrodes.
The dot itself is modeled as an
Anderson impurity center with Hamiltonian
\begin{equation}
\bbox{H}_{\rm dot}= \epsilon_0\sum_{\sigma }C^{\dagger}_{\sigma }C_{\sigma }+
UC^{\dagger}_{\uparrow }C_{\uparrow} C^{\dagger}_{\downarrow}C_{\downarrow}, 
\label{dot}
\end{equation}
where $C^{\dagger}_{\sigma }$ 
and $C_{\sigma }$ are dot electron operators.
The impurity site energy $\epsilon_0$ (counted from the Fermi energy $\mu$) 
is assumed to be far below the Fermi level $\epsilon_0<0$. The presence of a
strong Coulomb repulsion $U> -\epsilon_0$ 
between electrons in the same orbital
guarantees that the dot is at most singly occupied. 

Electron tunneling through the dot is described 
by the term,
\begin{equation}
\bbox{H}_{\rm t} = {\cal
T}\sum_{j=L,R}\sum_{\sigma}\Psi^{\dagger}_{j \sigma}(0)C_{\sigma} + {\rm h.c.},
\label{int} \end{equation} 
where ${\cal T}$ is an effective transfer amplitude.

The dynamics of the system is completely contained within
the evolution operator on the Keldysh contour $K$ \cite{Keldysh} which
consists of forward and backward oriented time paths. Its
kernel $J$ is given by a path integral, 
\begin{equation}
J=\int {\cal D} \bar{\Psi}{\cal D}\Psi {\cal D}\bar{C}{\cal D}C\exp(iS),
\label{pathint}
\end{equation}
over Grassman fields corresponding to the fermion operators, with
$\bar{\Psi} = (\Psi_{L \uparrow}^{\dagger},
\Psi_{L \downarrow}^{\dagger},\Psi_{R \uparrow}^{\dagger},
\Psi_{R \downarrow}^{\dagger})$ with obvious
definitions for $\Psi$, $\bar{C}$ and $C$. Moreover,
$S=\int_{K}Ldt$ is the action and $L$ is the 
Lagrangian pertaining to the Hamiltonian (\ref{tot}). 

In order to avoid dealing with fields defined on both branches
of the Keldysh contour one performs a rotation 
$C \to c$ and $\Psi \to \psi$ in Keldysh space: 
\begin{eqnarray}
\bar{c}=\bar{C}\sigma_z\hat Q^{-1}, \;\;\; c=\hat QC;\;\;\;
\hat Q&=\frac{1}{\sqrt{2}}&\left ( \begin{array}{cc}
1&-1\\
1&1
\end{array}\right) 
\label{transform}
\end{eqnarray}  
and similarly for $\bar{\psi}$ and $\psi$. Here $\sigma_z$ is 
the third Pauli matrix operating
in Keldysh space. 
The new Grassman variables $\bar{c}$, $c$, $\bar{\psi}$,
$\psi$ are now defined solely on the forward time branch. 
Averages of the corresponding products
of these fields determine the standard $2 \times 2$ 
Green-Keldysh matrix
\cite{Keldysh} composed of retarded ($\hat G^R$), advanced ($\hat G^A$) and
Keldysh ($\hat G^K$) Green functions which, in turn,
 are $2 \times2 $ matrices in
spin (Nambu) space. 

The path integral (\ref{pathint}) is expressed in terms of the new Grassman 
variables in the same way, and the action $S$ is now defined as
$S=S_{\rm dot}+S_0[\bar{\psi},\psi]$, where
\begin{equation}
S_{\rm dot}=\int dt\left[\bar{c}\left(i\frac{\partial}{\partial t}-\tilde{\epsilon}
\tau_{z}\right)c+\frac{U}{2}(\bar{c}c)^2\right], 
\label{sdot}
\end{equation}
\begin{eqnarray}
S_0=\int dt \sum_{j=L,R}\bigg[\int_jd{\bf r}\bar{\psi}_{j}
({\bf r},t)
\hat G_{j}^{-1}\psi_{j}({\bf r},t) \nonumber \\ 
+({\cal T}\bar{\psi}_{j}(0,t)\tau_{z}c(t)+{\rm c.c.})\bigg],
\label{SS}
\end{eqnarray}
where $\tilde{\epsilon}=\epsilon_{0}+U/2$ and the 
Pauli matrices $\tau_{x,y,z}$ act in Nambu space. 
The operator $\hat G_{L,R}^{-1}$ has the standard form 
\begin{equation} 
\hat G_{L,R}^{-1}(\xi )=i\frac{\partial}{\partial t}-
\tau_{z}\xi({\bf \nabla})+\tau_+\Delta_{L,R}+\tau_-\Delta^*_{L,R},
\label{G-1}
\end{equation}
where $\tau_{\pm}=(\tau_x\pm i\tau_y)/2$ and $\Delta_{L,R}$ are the (spatially
constant) BCS order parameters of the electrodes.

{\it Effective action and transport current.} The basic algorithm of our 
approach is to integrate
out the electron variables in the
superconducting electrodes which play
the role of an effective environment for the dot. This procedure yields
the influence functional $F[\bar{c},c]$ for the $c$-fields in the dot:
\begin{equation} 
F\equiv \exp (iS_{\rm env}[\bar{c},c])=
\int {\cal D} \bar{\psi}{\cal D}\psi 
\exp(iS_0[\bar{\psi},\psi]),
\label{pathint2}
\end{equation}
which is evaluated exactly.  Gaussian 
integration in (\ref{pathint2}) is carried out
separately for $L$- and $R$-electrodes.

Let us consider, say, the left superconductor and omit
the subscript $j=L$ for the moment.
The first step is to integrate 
out the fermion fields {\em inside} the superconductor
thereby arriving at an intermediate effective action in terms of the
fermion fields defined on the surface $x=0$.
It is useful at this point to Fourier transform
the fields $\psi (x,y)$ along the (translationally invariant)
$y$ direction. 
The problem then reduces to a one dimensional one 
with fermion fields $\psi_{k}(x)$ where $k$ is the quasiparticle
momentum in the direction normal to $x$.
In order to evaluate the Gaussian integral we will look for 
a saddle point field $\tilde \psi_k (x)$ defined by
$\hat G^{-1}(\xi_x) \tilde \psi_{k}(x)=0,$
where $\xi_x= -(1/2m)(\partial^2/\partial x^2)-\mu_k$ and $\mu_k=\mu -k^2/2m$. 

Decomposing $\tilde \psi$ into bulk and surface fields
$\tilde \psi_{k}(x)=\psi^{b}_{k}(x)+\psi_{k}(0)$ and integrating out
$\psi^{b}_{k}(x)$ we arrive at the intermediate effective action $\tilde S$ 
of a superconductor lead expressed only via the $\psi$-fields at the surface,
\begin {equation}
\tilde S=i\int dt \int dt'\sum_{k}\frac {v_{x}}{2}\bar{\psi}_{k}(0,t)
 \tau_{z}\hat g(t,t')\psi_{k}(0,t').
\label{SL}
\end{equation}
Here $v_x=\sqrt{2 \mu_{k}/m}$ and  
\begin{equation}
\hat g(t,t')=e^{\frac{i\varphi(t)\tau_z}{2}}\int \hat g(\epsilon
)e^{-i\epsilon (t-t')}
\frac{d\epsilon}{2\pi}e^{-\frac{i\varphi(t')\tau_z}{2}},
\label{gL1}
\end{equation}
is the Green-Keldysh-Eilenberger matrix of the (left) superconducting electrode  
\begin{eqnarray}
\hat g&=&\left ( \begin{array}{cc}
\hat g^R& \hat g^K\\
\hat 0&\hat g^A
\end{array}\right) , 
\label{g}
\end{eqnarray} 
$\varphi (t)=\varphi_0 
+2e\int^tV(t_1)dt_1$ is the time-dependent phase of
the superconducting order parameter and $V(t)$ is the electric potential of
the electrode. The Fourier transformed retarded and advanced 
Eilenberger functions have the standard form 
\begin{equation}
\hat g^{R/A}(\epsilon )= \frac{(\epsilon \pm i0)\tau_z +i|\Delta|\tau_y}
{\sqrt{(\epsilon \pm i0)^2-|\Delta|^2}},
\label{gL2}
\end{equation}
and
$\hat g^K=(\hat g^R-\hat g^A)\tanh (\epsilon /2T)$
is the Keldysh function.

The second step in our derivation 
amounts to integrating out the $\psi$-fields
on the surface. The integral
\begin{equation} 
\int {\cal D} \bar{\psi} (0){\cal D} \psi (0) 
\exp (i\tilde S+i\int dt({\cal T}\bar{\psi_k}(0)\tau_zc +{\rm c.c.}))
\label{pathint3}
\end{equation}
can easily be evaluated. Carrying out exactly the same procedure
for the right electrode, making use of the identity $\hat g_{L,R}^2=1$ 
and adding up the results we obtain
\begin{equation}
S_{\rm env}=i\Gamma \int dt \int
dt'\bar{c}(t)\tau_z\hat g_+(t,t')c(t').
\label{Eq_Sb}
\end{equation}
Here and below we define $\Gamma = 4\sum_k {\cal T}^2/v_x$ and $\hat g_{\pm}=(\hat g_L\pm \hat g_R)/2$.

Eq. (\ref{Eq_Sb}) is one of our central results. 
It enables the expression of
the kernel $J$ (\ref{pathint}) solely in terms of the fields $\bar{c}$ and
$c$: 
\begin{equation}
J=\int {\cal D} \bar{c}{\cal D}c \exp(iS_{\rm dot}+iS_{\rm env}),
\label{Jeff}
\end{equation}
where $S_{\rm dot}+S_{\rm env} \equiv S_{\rm eff}[\bar{c},c]$ represents the 
effective action for a quantum dot between two superconductors.

In order to complete our derivation let us express the current through
the dot in terms of the correlation function for the variables
$\bar{c}$ and $c$. Starting from the general expression for the current 
and representing the correlator for the $\psi$-fields in terms of that for
the $c$-fields we find,
\begin{equation}
I= \frac{e\Gamma}{2}{\rm Tr}[\hat g_\langle \bar cc\rangle |_K
+{\rm h.c.}].
\label{tok}
\end{equation}
Thus, the problem of calculating
the current through an interacting quantum dot
is reduced to that of finding the correlator 
$\langle \bar{c}c\rangle$ in the model defined by the effective
action $S_{\rm eff}=S_{\rm dot}+S_{\rm env}$ (\ref{sdot}), (\ref{Eq_Sb}). 
It should be emphasized that our approach is appropriate for
studying both equilibrium and nonequilibrium electron transport. 
In the noninteracting limit $U \to 0$ the results of previous studies
can be easily recovered within our formalism.


{\it Mean field approximation.}
Consider now the case $U \neq 0$ and 
decouple the interacting term in (\ref{sdot}) by means
of Hubbard-Stratonovich transformation \cite{Arovas,AG}
introducing additional scalar fields $\gamma_{\pm}$. The kernel $J$
now reads,
\begin{equation}
J=\int {\cal D} \bar{c} {\cal D} c {\cal D} \gamma_{+} {\cal D}\gamma_{-} 
\exp (iS[\gamma ]+iS_{\rm eff}|_{U=0}),
\label{HS}
\end{equation}
\begin{equation}
S[\gamma ]=\int dt\left(\bar{c}\gamma_{+}\sigma_{x}c+\bar{c}\gamma_{-}c 
-\frac{2}{U}\gamma_{+}\gamma_{-}\right).
\label{Seff1}
\end{equation}
Here we will assume that the effective Kondo temperature \cite{glaz}
$T_{K}$ =$\sqrt{U\Gamma}\exp{[-\pi|\epsilon_{0}|/2\Gamma]}$
is smaller than the superconducting gap $\Delta$. In this case 
interactions can be accounted for within
the mean field approximation. The fields 
$\gamma_{\pm}$ in (\ref{Seff1}) are considered as time-independent 
parameters determined self-consistently from the saddle point conditions
$\delta J/\delta \gamma_{\pm}=0$:
\begin{equation}
\gamma_{+}=\frac{U}{2}\int dt<\bar{c}c>, \;\;\;\; 
\gamma_{-}=\frac{U}{2}\int dt<\bar{c}\sigma_{x}c>.
\label{Eq_gamma}
\end{equation}
As it turned out from our numerical analysis the effect of the parameter
$\gamma_+$ is merely the renormalization of the tunneling rate $\Gamma$. 
Absorbing $\gamma_+$-terms in $\Gamma$ we arrive at
the final effective action of our model
\begin{eqnarray}
S_{\rm eff} [\gamma ]&=& \int \frac{d\epsilon}{2\pi} \int
d\epsilon'\bar{c} \hat M(\epsilon,\epsilon')c,\\
\hat M(\epsilon,\epsilon')&=&\delta
(\epsilon-\epsilon')(\epsilon +\gamma_{-}-\tau_z\tilde{\epsilon})+
i\tau_z\Gamma \hat g_{+}(\epsilon,\epsilon'). 
\label{Seff2}
\end{eqnarray}
 
\begin{figure}[htb]
\includegraphics[width=0.45\textwidth,keepaspectratio]{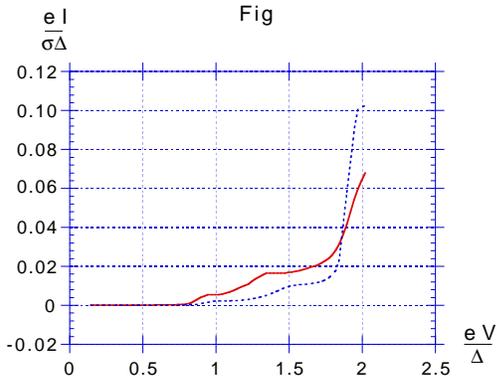}
\caption
{The I-V curves for an $SAS$ junction at subgap voltages.
The parameters are $\epsilon_0 =-1.5$,  $\Gamma=0.6$, $U=2.4$ (solid
curve) and 2.7 (dashed curve).}
\label{fig1}
\end{figure}
{\it Subgap current in SAS junctions}. In order to find the 
correlator $\langle \bar cc\rangle =i\hat M^{-1}$
and the current (\ref{tok}) we numerically inverted the
matrix (\ref{Seff2}) and simultaneously solved the self-consistency 
equation for $\gamma_-$ (\ref{Eq_gamma}). The resulting $I-V$ characteristics 
for an $SAS$ junction in the presence of Coulomb interaction are 
displayed in Fig.1. One observes all the main features predicted
within our simple picture of an interplay between MAR and Coulomb
interaction: (i) at relatively low voltages $V \leq V_{th}$ 
MAR current is essentially suppressed due to interaction, (ii) for 
higher voltages 
(but still $eV < 2\Delta$) MAR is possible and results in a nonzero subgap
current which increases with $V$ and (iii) the subharmonic peaks
in the differential conductance occur and are shifted to higher voltages as  
compared to the noninteracting case. 
An increase of $U$ results in a stronger current suppression and a 
more pronounced shift of the subharmonic peaks. 
Close to the gap edge $eV=2 \Delta$ the current shoots up sharply.

The parameters used in our numerical analysis are chosen in a way
to observe all the key features (i), (ii) and (iii). It is interesting
to quantitatively compare the results presented in Fig. 1 with the predictions
(\ref{n})-(\ref{peaksEc}) of an oversimplified ``$E_C$-based''
model. Let us estimate the effective value of $E_C$ (which 
is, strictly speaking, a function of $U$, $\epsilon_0$ $\Gamma$ and $V$ in 
our calculation) with the aid of eq. (\ref{threshold}) and the $I-V$
curves of Fig. 1. We find $E_C \approx 0.2\Delta$ for $U=2.4\Delta$
and $E_C \approx 0.25\Delta$ for $U=2.7\Delta$. Obviously, these
values of $E_C$ are smaller than $2\Delta /3$ and, hence, a finite 
subgap conductance is expected at $eV\gtrsim \sqrt{2\Delta E_C} \gg E_C$. This 
is precisely what we observe in Fig. 1. For such values of $V$ and $E_C$
eq. (\ref{n}) yields $n \leq 3$, i.e. only two subharmonic peaks 
(with $n=2,3$) can occur. This is exactly the case in Fig. 1. Finally, 
substituting the above  
values of $E_C$ into eq. (\ref{peaksEc}) we can estimate the magnitudes
of the peak shifts $\delta V_n$. For
$U=2.4\Delta$ we find $\delta V_n \sim 0.3\Delta$ for $n=2$ and
$\sim 0.26 \Delta$ for $n=3$. Analogous values for $U=2.7\Delta$ are
respectively $\sim 0.38\Delta$ and $\sim 0.33 \Delta$. These values are in 
a reasonably good agreement with our numerical results.

In summary, we presented a detailed analysis of an SAS junction 
at finite bias and derived its effective action
using Keldysh path-integral techniques. Our approach
applies for both equilibrium and nonequilibrium current transport
in the presence of interactions. The repulsive Coulomb interaction 
leads to novel effects in the pattern of the subgap current.
In particular, it shifts the peaks of the differential conductance 
toward larger bias. When the interaction is 
sufficiently strong the subgap current is highly suppressed. 
Our theoretical predictions can be directly tested in experiments
with superconducting quantum dots.

We acknowledge useful discussions with J.C. Cuevas and J. von Delft. 
This research is supported by DIP German Israel Cooperation project, 
by the Israeli Science Foundation grant {\em Center of Excellence} and 
by the US-Israel BSF.

\end{multicols}
\end{document}